\documentclass[sigconf,nonacm]{acmart}

\usepackage{subcaption}
\usepackage{placeins}
\usepackage{enumitem}
\setlist{noitemsep, topsep=2pt, parsep=0pt, partopsep=0pt}
\begin{document}

\title{A Multi-Dimensional Score for Detecting Lesser-Known Points of Interest in Mountain Regions}

\author{Adriana Pompejano}
\affiliation{%
  \institution{Universite de Pau et des Pays de l’Adour, E2S UPPA, LIUPPA,}
  \city{Anglet}
  \country{France}
}
\email{adriana.pompejano@univ-pau.fr}

\author{Sébastien Laborie}
\affiliation{%
  \institution{Universite de Pau et des Pays de l’Adour, E2S UPPA, LIUPPA,}
  \city{Bayonne}
  \country{France}
}
\email{sebastien.laborie@univ-pau.fr}

\author{Maxime Masson}
\affiliation{%
  \institution{Universite de Pau et des Pays de l’Adour, E2S UPPA, LIUPPA,}
  \city{Anglet}
  \country{France}
}
\email{maxime.masson@univ-pau.fr}

\author{Christian Sallaberry}
\affiliation{%
  \institution{Universite de Pau et des Pays de l’Adour, E2S UPPA, LIUPPA,}
  \city{Pau}
  \country{France}
}
\email{christian.sallaberry@univ-pau.fr}

\author{Philippe Roose}
\affiliation{%
  \institution{Universite de Pau et des Pays de l’Adour, E2S UPPA, LIUPPA,}
  \city{Anglet}
  \country{France}
}
\email{philippe.roose@univ-pau.fr}

\renewcommand{\shortauthors}{Pompejano et al.}

\begin{abstract}
Overtourism concentrates visitor demand on few iconic sites, while other points of interest (POIs) in the same area often remain known only to a narrower public. This imbalance is especially critical in mountain environments, where limited infrastructure amplifies environmental pressures. Discovering lesser-known POIs is relevant for redistributing tourism demand spatially, valuing underrepresented heritage, and temporally, easing high seasonality pressures on popular POIs.
Across the literature, definitions of lesser-known POIs or hidden gems remain scattered, and existing scoring approaches remain incomplete, relying on narrow indicators (e.g., number of photos, reviews, likes) which fail to characterize mountain POIs holistically. These approaches are further
constrained by their reliance on commercial proprietary sources, or by questionnaires that are subject to memory bias. To address these gaps, we introduce a novel multi-dimensional metric, the Obscurity Score, that groups fragmented literature into four key dimensions (\emph{International Awareness, Attractiveness, Promotion, and Accessibility}) to systematically detect and characterize lesser-known POIs and enable their promotion as a first step toward mitigating overtourism. To operationalize this score, we develop a generic pipeline mapping open data sources onto these four dimensions. Preliminary results from a web-based prototype demonstrate the practical relevance of the proposed approach.

\end{abstract}


\maketitle

\section{Introduction}

In recent years, the tourism sector has experienced significant growth, a phenomenon that
generates substantial benefits for local economies but also has a considerable impact on
ecosystems. Despite this increase,
tourism tends to concentrate on a limited number of high-visibility destinations, while
equally meaningful places nearby remain unknown to the wider public
\cite{bertocchi2021tourism,rogowski2025managing}. This asymmetry has two faces: the overcrowding of iconic attractions, and the invisibility of other sites that remain absent
from mainstream platforms \cite{pradipta2021journee,rajaonarivo2023few}. That imbalance is particularly critical in mountain environments, where limited infrastructure, difficult
emergency services and ecological fragility amplify the impact of visitor concentration
\cite{senetra2020evaluation}. Detecting lesser-known Points of Interest (POIs) is therefore relevant both for
the spatial dispersion of tourism demand and for the promotion of territorial heritage,
benefiting both residents and tourists.

Current recommendation platforms are not yet designed to support lesser-known POIs, as they prioritize popularity over the discovery of hidden gems. In mountain contexts, for example, popular outdoor navigation applications such as AllTrails\footnote{\url{https://www.alltrails.com}}, or Visorando\footnote{\url{https://www.visorando.com}} assist hikers in identifying nearby routes and activities. However, these platforms typically rely on collaborative filtering algorithms that recommend locations based on high engagement and top user ratings. This logic produces a feedback loop, directing visitors toward already saturated sites while keeping potentially valuable alternatives invisible. As an alternative, hikers increasingly turn to generative AI tools to plan itineraries and discover lesser-known routes, but this practice is neither reliable \footnote{\url{https://www.digitaltrends.com/computing/chatgpts-hiking-advice-left-two-hikers-stranded-on-a-mountain-in-poland/}} nor safe \footnote{\url{https://www.tourisme-haute-savoie.com/en/sports-and-leisure/hiking/mountain-hiking-why-trusting-ai-to-prepare-your-itinerary-is-a-dangerous-trap/}}.

Prior studies that explicitly target lesser-known POIs face three recurring limitations. 

First, they mostly characterize POI popularity from a single perspective, often relying on individual indicators such as photo counts, distance, or review volume \cite{roy2022information,lin2020less,zhuang2014anaba,buranasing2021recommendation,liu2025assessing}. Even studies that combine multiple indicators \cite{krsak2026geosite} focus exclusively on online visibility and partially on accessibility, failing to consider local recognition and promotional efforts.

Second, most methods reduce recognition to a binary state (popular vs. hidden) \cite{pradipta2021journee,buranasing2021recommendation,rajaonarivo2023fewshot}. Even continuous rankings \cite{zhuang2014anaba,krsak2026geosite} fail to capture intermediate tiers (\emph{known} and \emph{partially known}) between overcrowded (\emph{mainstream}) and hidden (\emph{lesser-known}) sites. This distinction matters operationally: for instance, Destination Management Organizations (DMOs) need to know how much promotional margin each POI still has, none for \emph{mainstream} sites, little for \emph{known} sites nearing saturation, moderate for \emph{partially known} ones, and full for \emph{lesser-known} sites.

Third, existing approaches depend on paid or proprietary databases such as X\footnote{\url{https://x.com}}, Flickr\footnote{\url{https://www.flickr.com}}, Tripadvisor\footnote{\url{https://www.tripadvisor.com}} or Google Maps reviews\footnote{\url{https://maps.google.com}} \cite{rajaonarivo2023fewshot,zhuang2014anaba,krsak2026geosite}, which attract only a specific type of user, or on questionnaires that are subject to respondent memory bias \cite{lin2020less}. Open sources offer a viable alternative, especially in mountain contexts, where Volunteered Geographic Information (VGI) benefits from an active community
of outdoor contributors that systematically documents trails and POIs \cite{vukalic2026openstreetmap}, offering more consistent spatial coverage than social-platform check-ins.

To address these limitations, we propose a multi-dimensional characterization of mountain POIs. We synthesize prior work into four complementary dimensions that emerge from the literature. \emph{International Awareness} captures how widely a site is recognized beyond its local context; \emph{Attractiveness} reflects the extent of user-generated photos available for a POI; \emph{Promotion} measures the degree to which a site is actively marketed through official channels; \emph{Accessibility} accounts for how easily a site can be physically reached. While previous research on lesser-known POIs examines such factors in isolation, our work integrates the multi-dimensional aspect into a unified score that reflects how well-known or lesser-known a site is.
This score is computed through a generic pipeline that combines indicators across these dimensions, using exclusively open data, to calculate how lesser-known a site is. We define this continuous property as \emph{obscurity}, adopting the term from Zhuang et al. \cite{zhuang2014anaba}, who use it to describe sightseeing spots that are less well-known yet still worth visiting. This aligns with literature that measures site recognition on a scale, such as Lin et al. \cite{lin2020less} using \emph{familiarity} or Kršák et al. \cite{krsak2026geosite} using \emph{digital invisibility}.

Our central research question is whether an Obscurity Score derived entirely from open spatio-temporal
data can effectively distinguish lesser-known mountain POIs from popular ones, while also
capturing intermediate levels of site recognition.

The main contributions of this paper are threefold: 

(i) a four-dimensional Obscurity Score that characterizes POI recognition, mapping scores onto a four-tier ordinal scale (\emph{mainstream, known, partially known, lesser-known}), where higher values indicate greater obscurity (\emph{lesser-known POIs}) and lower values correspond to \emph{mainstream} attraction points; 

(ii) a federation of open data from multiple sources to align it with the four dimensions of our score;

(iii) a visual platform that allows users to observe a gradation of place types (from \emph{mainstream} to \emph{lesser-known}) within a given territory. This tool is based on a generic pipeline that integrates, filters, and allows customization of the Obscurity Score calculation. Absent in existing literature, this decision-support tool empowers stakeholders (such as DMOs) to make informed choices, such as developing new tourism offerings or redistributing visitor flows.

This paper is organized as follows: Section~\ref{sec:related} reviews related work on POI data sources and lesser-known POI characterization; Section~\ref{sec:dimensions} justifies our multi-dimensional characterization; Section~\ref{sec:score} introduces the Obscurity Score; Section~\ref{sec:pipeline} details our generic pipeline; Section~\ref{sec:usecase} presents the visual platform and preliminary case study; Section~\ref{sec:discussion} discusses preliminary results; and Section~\ref{sec:conclusion} concludes with future work.

\section{Related Work}
\label{sec:related}
This section covers POI data sources (Section~\ref{sec:poi-sources}) and how lesser-known POIs have been characterized (Section~\ref{sec:related-lk}).

\subsection{POI Sources and Volunteered Geographic Information Quality}
\label{sec:poi-sources}

A POI is commonly described as a set of coordinates associated with an identifier, extensible to a richer system comprising multilingual names, categories, and addresses \cite{ogc2024poi}. Milias and Psyllidis \cite{milias2021assessing} refer to POIs as digital representations of real-world amenities. 

Among spatial data sources, two primary families coexist: paid-based sources and open datasets. Commercial proprietary datasets (e.g., Google Places, Foursquare\footnote{\url{https://foursquare.com}}, AMap\footnote{\url{https://www.amap.com}}) offer high positional precision and verified business records. However, their licensing terms preclude long-term storage, redistribution, and independent reproducibility. Additionally, they are often sold as static snapshots without temporal updates. Conversely, open sources rely on volunteered geographic information, a collaborative mapping system conceptualized by Goodchild \cite{goodchild2007citizens} and operationalized today through OpenStreetMap (OSM). OSM is openly licensed and continuously updated by a distributed community of contributors. A recognized limitation of Volunteered Geographic Information (VGI) is that completeness and accuracy depend on the size and engagement of the local contributor community rather than centralized quality control. Recent evidence, however, qualifies this concern for mountain settings: while VGI was historically concentrated in urban areas \cite{haklay2010good}, mountain regions have progressively achieved high-quality coverage thanks to sustained contributions from hiking and alpine communities \cite{vukalic2026openstreetmap}.

Two further properties motivate our reliance on open sources. First, lesser-known sites are precisely those least likely to appear in commercially curated directories. Second, every OSM edit is timestamped, enabling newly mapped or modified POIs to be detected automatically as the underlying dataset evolves.

\subsection{Characterizing Lesser-Known POIs}
\label{sec:related-lk}

The literature converges on describing lesser-known POIs or hidden gems as places characterized by low external visibility and scarce information \cite{rajaonarivo2023few,pradipta2021journee}. Beyond this shared core, definitions diverge in their emphasis: some highlight a lack of marketing and promotion \cite{buranasing2021recommendation,lin2020less,pradipta2021journee}, others stress their significance for local culture \cite{rajaonarivo2023few}. 
This conceptual heterogeneity leads to varied operationalizations: for instance, Rajaonarivo and Mine \cite{rajaonarivo2023fewshot} set fixed quantitative thresholds, requiring fewer than 500,000 Google results and under 20 comments or photos on major platforms. Beyond this approach, the indicators used across prior studies fall into four recurring families.

First, places of broad interest tend to be documented in several languages, while those of only local relevance appear in few or none \cite{roy2022information}. Second, user-contributed photos serve as a reliable proxy for revealed attractiveness (i.e., the interest expressed through documentation rather than intrinsic aesthetic quality) and actual site visitation \cite{krsak2026geosite,lin2020less,zhuang2014anaba}. Third, dedicated promotional channels and social media managed by destination organizations (DMOs) act as direct indicators of commercially marketed sites \cite{buranasing2021recommendation,hervas2026social}. Fourth, physical accessibility influences visitation levels, given that easily reachable locations are more likely to become tourist hotspots \cite{liu2025assessing,krsak2026geosite}; academic studies specific to mountain environments narrow the focus further, concentrating on topography. For example, they highlight that elevation and steepness predict hiking suitability \cite{naimi2024machine}, while difficulty determines trail classifications \cite{krevs2023reconsidering}.

Yet existing studies evaluate these indicators in isolation or rely on rigid statistical weighting like Principal Component Analysis (PCA) and entropy \cite{lin2020less,krsak2026geosite}, without adapting to context. Mountain settings vary widely, and the relative importance of each indicator can shift accordingly, calling for a unified, parameterizable score that integrates them and adapts their weights to local conditions.

\section{Methodology: a novel approach to detect lesser-known POIs}

This section details our methodology: we first introduce and justify the four core dimensions essential to analyze a location (Section~\ref{sec:dimensions}), define a flexible multi-dimensional Obscurity Score (Section~\ref{sec:score}), and illustrate a generic open-data pipeline aligned with the dimensions, allowing us to experiment with our proposal (Section~\ref{sec:pipeline}).

\subsection{A multi-dimensional analysis of POIs}
\label{sec:dimensions}

While the indicators reviewed in Section~\ref{sec:related-lk} have each been used in prior work, their grouping into four complementary dimensions and their joint integration into a single Obscurity score are contributions of this paper. We introduce four dimensions for analyzing lesser-known POIs, one per indicator family: \emph{International Awareness, Attractiveness, Promotion, and Accessibility}. We treat these four dimensions as a starting hypothesis, open to refinement or extension as our experiments progress. This section justifies why they are jointly necessary before combining them into a unified score. To our knowledge, no existing approach integrates all four jointly: Kršák et al. \cite{krsak2026geosite}, the only reviewed framework combining multiple indicators, covers only \emph{Attractiveness} (photo volume) and, partially, \emph{Accessibility} (road distance from a city center), leaving \emph{International Awareness} and \emph{Promotion} unaddressed.

Each POI, denoted $p$, is evaluated across the four dimensions $D_1$--$D_4$, each operationalized through observable indicators summarized in Table~\ref{tab:dimensions}.

\begin{table}[htbp]
\centering
  \caption{Overview of the four dimensions included in the Obscurity Score. Weaker signals (for $D_4$, greater effort) indicate higher obscurity.}
  \label{tab:dimensions}
  \vspace{-6pt}
  \small
  \renewcommand{\arraystretch}{1.1}
  \setlength{\tabcolsep}{4pt}
  \begin{tabular}{@{}p{0.34\linewidth} p{0.62\linewidth}@{}}
    \toprule
    \textbf{Dimension} & \textbf{Indicators} \\
    \midrule
    \textbf{D1} International Awareness & Multilingual coverage and public search/view volume \\
    \textbf{D2} Attractiveness & Volume of media assets (photos) and documentation history \\
    \textbf{D3} Promotion & Inclusion in structured itineraries and dedicated marketing media \\
    \textbf{D4} Accessibility & Physical effort and travel time required from key transit nodes (e.g., parking lots, routes) \\
    \bottomrule
  \end{tabular}
  \vspace{-8pt}
\end{table}

Each dimension addresses a gap the others cannot fill. \emph{International Awareness} distinguishes a site known only regionally from one recognized worldwide, even when \emph{Attractiveness} or \emph{Accessibility} are identical. \emph{Promotion} separates organic popularity from active marketing: institutional efforts such as an official POI website signal an intent to increase visibility that \emph{Attractiveness} and \emph{International Awareness} cannot capture on their own. \emph{Accessibility}, in turn, distinguishes a genuinely obscure but remote site from one that is merely hard to reach despite heavy promotion. Finally, \emph{Attractiveness}, understood here as revealed rather than intrinsic appeal, is the most widely used indicator in the literature \cite{zhuang2014anaba,lin2020less,krsak2026geosite}, and omitting it would discard the dimension with the broadest empirical grounding.
The concrete computation of each indicator from open data sources is illustrated with a practical example in Section~\ref{sec:usecase}.

\subsection{The Obscurity Score}
\label{sec:score}

Building on the four dimensions introduced above, we capture how lesser-known a site is through a continuous property we call \emph{obscurity}, following the terminology of Zhuang et al. \cite{zhuang2014anaba} and consistent with related scales in the literature, such as \emph{familiarity} in Lin et al. \cite{lin2020less} or \emph{digital invisibility} in Kršák et al. \cite{krsak2026geosite}. Unlike these approaches, we capture obscurity through a parameterizable score, whose relative weights can be adjusted to reflect how the importance of each dimension varies across mountain contexts.

The Obscurity Score $S(p)$ is a composite score that aggregates the four dimensions defined in Section~\ref{sec:dimensions}:
\setlength{\abovedisplayskip}{3pt}
\setlength{\belowdisplayskip}{3pt}
\begin{equation}
  S(p) = w_1 D_1(p) + w_2 D_2(p) + w_3 D_3(p) + w_4 D_4(p),
  \label{eq:score}
\end{equation}
where $D_1(p),\ldots,D_4(p) \in [0,1]$ are the normalized dimension values of POI's ($p$) and $w_1,\ldots,w_4$ are the dimension weights, constrained to sum to one. Each dimension $D_i(p) \in [0,1]$ is normalized to highlight obscurity: values close to 1 indicate weak international documentation, lack of promotion, sparse media coverage, or difficulty of access, while values close to 0 indicate high internationality, strong attractiveness and promotion, or easy accessibility.

The weights will be determined through a two-stage process: an equal 0.25 baseline, followed by a grid search that will maximize agreement between the Obscurity Score and POI ratings from 4–5 local specialists per case study (e.g., alpine guides, hut managers, tourism professionals: people with hands-on knowledge of visitor flows), collected via a structured field questionnaire. We will adopt this calibration strategy, rather than a purely statistical scheme (e.g., PCA in \cite{krsak2026geosite}), to ground each dimension's weight directly in domain expertise.

Because Equation~\ref{eq:score} keeps per-dimension scores independent of the weights themselves, this modular design makes the framework flexible: alternative weight configurations can be explored without recalculating the underlying indicators, and the calibrated weights will remain an adaptable baseline rather than a fixed constant. We further hypothesize that optimal weights may vary across geographical contexts. Testing this variability across different mountain regions represents a primary direction for future validation.

The resulting continuous score $S(p)$ is classified into four ordinal tiers: \emph{mainstream}, \emph{known}, \emph{partially known}, and \emph{lesser-known}. In contrast to traditional binary classifications, this graduation is designed to approximate the size of the community that recognizes a place: ranging from attractions frequented by mass tourists, through sites regularly visited by the wider public but not yet overcrowded, to places visited mainly by regular hikers or local guides, down to sites with almost no recorded visitors at all. In the absence of calibrated weights, tier boundaries are set using Jenks natural breaks ($k{=}4$) \cite{jenks1967data} on the empirical distribution of $S(p)$ across POIs, rather than fixed intervals; we adopt this approach over equal intervals or quantiles because it groups scores based on natural clusters rather than assuming an even distribution. This is essential for our dataset, as the case study shows a clear asymmetry in score distribution (Section~\ref{sec:usecase}). Weights and boundaries will be calibrated via grid search on an expert gold standard (Fleiss' $\kappa$) and evaluated on experts excluded from calibration (F1-score).

\begin{figure}[htbp]
  \centering
  \includegraphics[width=\linewidth]{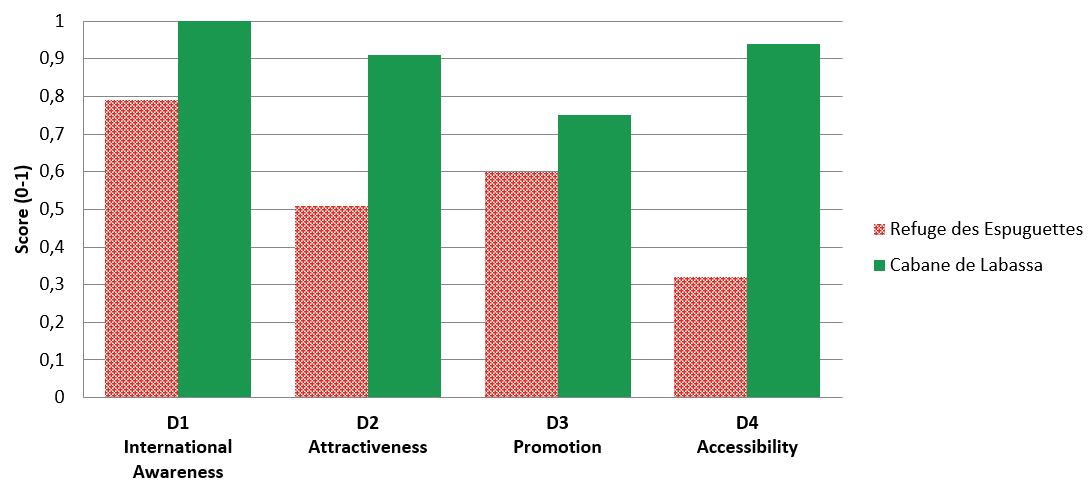}
  \caption{Comparison of dimension scores ($D_1(p)$–$D_4(p)$) for Refuge des Espuguettes and Cabane de Labassa.}
  \label{fig:hut-comparison}
  \vspace{-8pt}
\end{figure}

Figure~\ref{fig:hut-comparison} illustrates the score's application, under equal weighting, with two real huts $\sim$2.8~km apart near the Cirque de Gavarnie (Pyrenees, France), to demonstrate how the score separates sites of comparable proximity to the same landmark: Refuge des Espuguettes ($D_1(p_1){=}0.79$, $D_2(p_1){=}0.51$, $D_3(p_1){=}0.60$, $D_4(p_1){=}0.32$) scores $S(p_1){=}0.55$, classified as \emph{mainstream}, while Cabane de Labassa ($D_1(p_2){=}1.00$, $D_2(p_2){=}0.91$, $D_3(p_2){=}0.75$, $D_4(p_2){=}0.94$) scores $S(p_2){=}0.90$, classified as \emph{lesser-known}. The full mapping from raw open-data indicators to each $D(p)$ value is detailed in Section~\ref{sec:usecase}.

\subsection{Pipeline Overview}
\label{sec:pipeline}

Having defined the four dimensions and the Obscurity Score that aggregates them, we now propose a six-stage generic pipeline for identifying lesser-known sites within a region to support territorial analysis. The pipeline applies to any geographic region, accepts different data sources, adapts as underlying data evolve over time, and supports multiple languages.

\begin{description}
  \item[(A) Collection.] POIs, along with their spatial coordinates and descriptive tags, are retrieved from a VGI repository over a defined bounding area.
  \item[(B) Pre-filtering.] An exclusion rule is applied before scoring: Unnamed POIs without any photo are excluded entirely rather than down-weighted. This ensures that the score measures true obscurity rather than incomplete data mapping \cite{goodchild2007citizens}. Additionally, near-identical features within a fixed spatial radius are deduplicated. 
  \item[(C) Enrichment.] Each retained POI is linked to external open sources, such as a multilingual knowledge base or a media repository, to gather additional descriptive attributes.
  \item[(D) Scoring.] Four key dimensions (\emph{International Awareness}, \emph{Attractiveness}, \emph{Promotion}, and \emph{Accessibility}) are computed from underlying specific indicators. Each dimension is normalized to the unit interval $[0, 1]$, and they are weighted and aggregated into a single, unified Obscurity Score for each POI (detailed in Section~\ref{sec:score}).
  \item[(E) Ranking.] The continuous score is classified into four ordinal tiers, categorizing POIs along a gradient from \emph{mainstream} to \emph{lesser-known} (Section~\ref{sec:score}).
  \item[(F) Calibration and evaluation.] A grid search adjusts the weights of step (D) to match expert judgments (e.g., tourism professionals): the result is then evaluated on experts excluded from calibration (Section~\ref{sec:score}).
\end{description}

We implemented this pipeline as the web-based prototype presented next (Section~\ref{sec:usecase}), aiming to classify mountain POIs according to their obscurity level.

\section{Prototype and Preliminary Use Case}
\label{sec:usecase}

The pipeline is operationalized as a web-based prototype \footnote{\url{https://apompejano.perso.univ-pau.fr/poi-finder/}} in which a user selects an area of
interest and the resulting tiers are rendered cartographically. 

\textbf{Study area.} The chosen bounding box (42.677$^\circ$--42.739$^\circ$N,
$-$0.067$^\circ$--0.073$^\circ$E) spans the French-Spanish border in the Pyrenees, covering parts of the Parc National des Pyrénées and the Mont Perdu UNESCO World Heritage property. The area was selected for its diverse POI types (lakes, waterfalls, peaks, mountain huts) rather than popularity, and for its high density of eligible POIs, a realistic mix of mainstream and hidden sites. For instance, it features prominent landmarks such as the Cirque de Gavarnie, the Cascade de Gavarnie, and the Brèche de Roland.

\textbf{Sources.} In this case study, POIs and tags are retrieved from OpenStreetMap. The underlying indicators are calculated using multilingual coverage and stable identifiers from Wikidata alongside traffic statistics from Wikimedia Pageviews ($D_1(p)$), imagery from Wikimedia Commons ($D_2(p)$), official trail network labels from OpenStreetMap ($D_3(p)$), and walking times from Open Source Routing Machine ($D_4(p)$). For $D_4(p)$, distance is measured to the nearest transit node (a parking area, a marked-trail entry point, or a cable-car station), whichever is closest to the POI. Other VGI sources could have been used, but we relied on these for their consistent coverage across diverse mountain territories, allowing comparability across the future case studies.

\begin{figure}[t]
  \centering
  \includegraphics[width=\columnwidth]{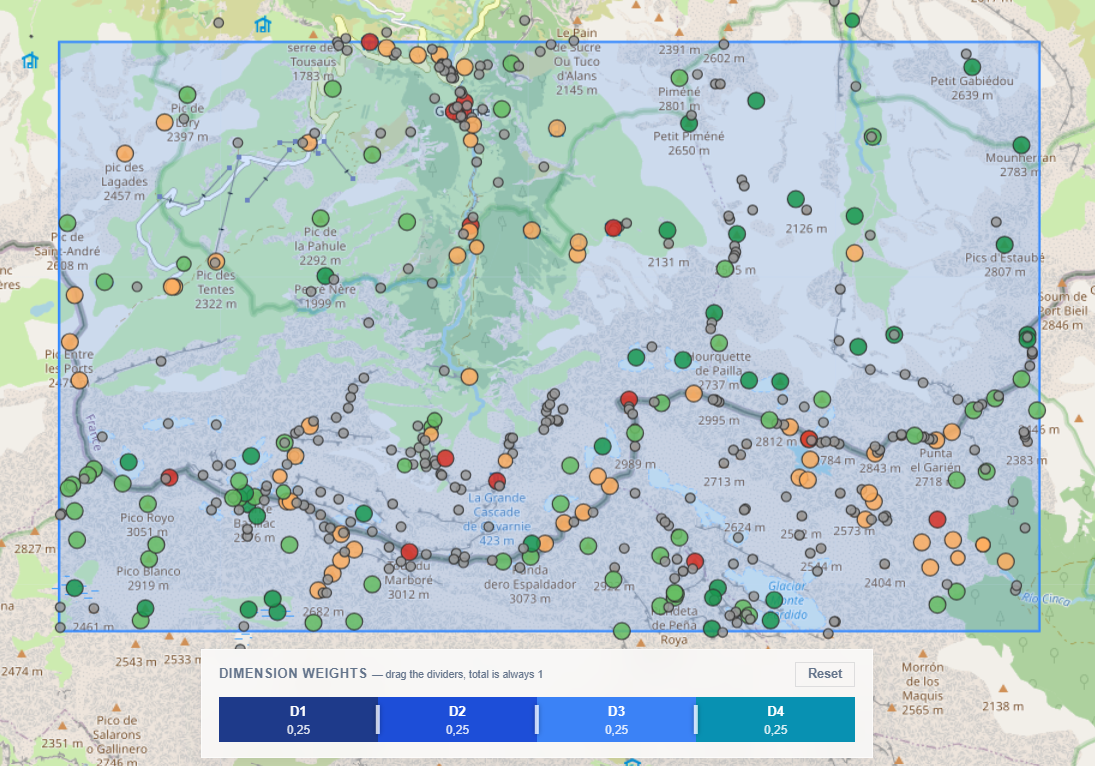}
  \caption{Color-classified Obscurity Score (red = \emph{mainstream}, orange = \emph{known}, green = \emph{partially known}, dark green = \emph{lesser-known}; grey = excluded); bottom panel shows current weighting.}
  \label{fig:map}
\end{figure}

\textbf{Dataset volume.} On this area, the pipeline retrieved 499 raw POIs. After pre-filtering and spatial deduplication (Section~\ref{sec:pipeline}, step B), 316 POIs were excluded and 183 were scored. Among the scored POIs, 41 were classified as \emph{lesser-known}, 67 as \emph{partially known}, 61 as \emph{known} and 14 as \emph{mainstream} (Figure~\ref{fig:map}).

\begin{figure}[t]
  \centering
  \begin{subfigure}[t]{0.48\columnwidth}
    \centering
    \includegraphics[width=\linewidth]{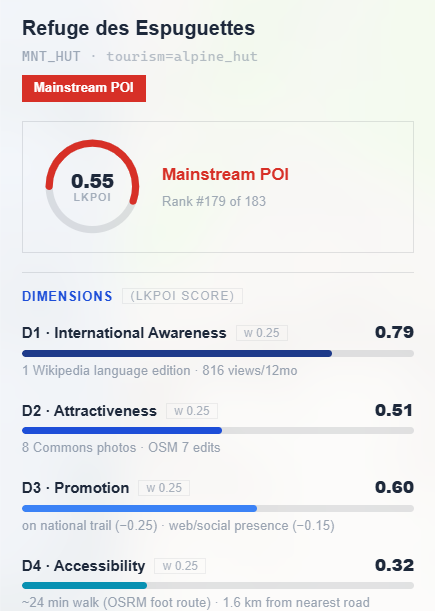}
    \caption{Refuge des Espuguettes: $D_1(p_1)=0.79$, $D_2(p_1)=0.51$, $D_3(p_1)=0.60$, $D_4(p_1)=0.32$, Obscurity Score $S(p_1)=0.55$}
    \label{fig:example-mainstream}
  \end{subfigure}
  \hfill
  \begin{subfigure}[t]{0.48\columnwidth}
    \centering
    \includegraphics[width=\linewidth]{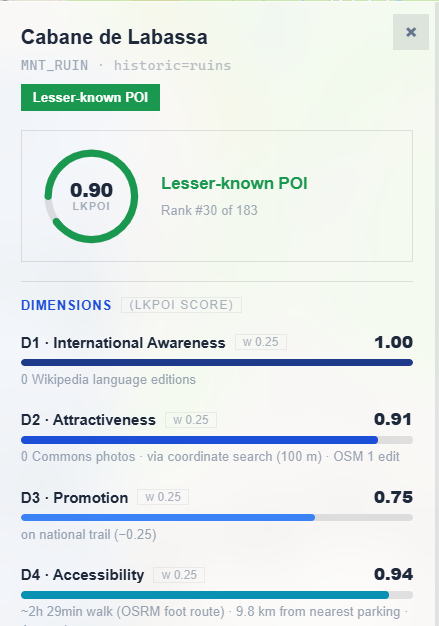}
    \caption{Cabane de Labassa (ruins): $D_1(p_2)=1.00$, $D_2(p_2)=0.91$, $D_3(p_2)=0.75$, $D_4(p_2)=0.94$, Obscurity Score $S(p_2)=0.90$}
    \label{fig:example-lk}
  \end{subfigure}
  \caption{Two scored POIs: an equipped (\emph{mainstream}) mountain hut on a
  national trail, and a rustic (\emph{lesser-known}) hut.}
  \label{fig:examples}
\end{figure}

\textbf{Illustrative examples.} Figure~\ref{fig:examples} details two POIs at opposite ends of the scale, with their $D_1(p)$--$D_4(p)$ values, classification and rank: Refuge des Espuguettes and Cabane de Labassa.

To illustrate the scoring process, we detail the computation of all four dimensions.

For $D_1(p)$, Refuge des Espuguettes has 1 Wikipedia language edition ($langScore=0.90$) and 68/month pageviews, falling in the $\leq200$ bin ($pageviewsScore=0.67$), giving $D_1(p_1)=0.5(0.90)+0.5(0.67)=0.79$; 
Cabane de Labassa has no Wikipedia article at all, so $langScore=1.00$ and the pageviews term is omitted, leaving $D_1(p_2)=1.00$. 

For $D_2(p)$, Espuguettes has 8 Commons photos ($photoScore=0.65$) and 7 OSM edits (mapping-maturity $=0.25$), giving $D_2(p_1)=0.65(0.65)+0.35(0.25)=0.51$; Labassa has 0 Commons photos ($photoScore=1.00$) and 1 OSM edit (mapping-maturity $=0.75$), giving $D_2(p_2)=0.65(1.00)+0.35(0.75)=0.91$. 

For $D_3(p)$, both huts sit on a national hiking route ($-0.25$); Espuguettes additionally has a website or social media presence ($-0.15$), giving $D_3(p_1)=1.0-0.25-0.15=0.60$, while Labassa has none, giving $D_3(p_2)=1.0-0.25=0.75$. 

For $D_4(p)$, walking time is mapped to a score via a smooth curve centered at 45~minutes\footnote{$D_4(t) = 0.05 + 0.90/(1+e^{-0.04(t-45)})$, where $t$ is the routed walking time in minutes.}: shorter walks score low (near 0.05), longer walks score high (near 0.95), with 45~minutes as the midpoint. Espuguettes, a 24-minute walk from the nearest road, falls well below this midpoint and scores $D_4(p_1)=0.32$; Labassa, a 2h29 walk from the nearest parking, falls far above it and scores $D_4(p_2)=0.94$.

The bin boundaries and sub-weights used above (e.g., the 0.65/0.35 split in $D_2(p)$, the 45-minute logistic midpoint in $D_4(p)$) are current design defaults rather than empirically derived constants; refining them against a larger expert-validated sample is left for future work.

\textbf{Validation protocol.} Identification of local respondents for this territory (alpine guides, hut managers, and tourism professionals) is currently ongoing. For each POI, respondents will rate visitor frequency on a four-point scale (mainstream to lesser-known); POIs unrecognized by all respondents will be flagged for exclusion.

\textbf{Data export and analytics.} Beyond the prototype itself, every POI score and the raw signal underlying it are exported as a single CSV per run, making it straightforward to aggregate or analyze
open-data evidence for future comparative studies without re-executing the pipeline.

\section{Discussion and preliminary results}
\label{sec:discussion}

The case study described above produced preliminary empirical findings. Table~\ref{tab:examples} illustrates the ranking obtained under equal dimension weighting (0.25) for three OSM-tagged categories in the Pyrenees case study: cirques, peaks and mountain huts. Each mainstream POI is paired with a nearby known, partially known, or lesser-known alternative, such as the popular Refuge des Espuguettes with the more rustic Cabane de Labassa (distance shown here for reference only). These preliminary results still need to be validated by local experts, with dimension weights calibrated to approach the resulting gold standard.

\begin{table*}[t]
  \caption{Same-category POI comparisons (Gavarnie case study, $w_i{=}0.25$).}
  \label{tab:examples}
  \small
  \renewcommand{\arraystretch}{1.05}
  \setlength{\tabcolsep}{4pt}
  \begin{tabular}{@{}l l l c r r r r l r@{}}
    \toprule
    \textbf{Category} & \textbf{POI} & \textbf{Class} & $\boldsymbol{S(p)}$ & \textbf{Wiki langs} & \textbf{Views/12mo} & \textbf{Photos} & \textbf{OSM edits} & \textbf{Promotion} & \textbf{Walk} \\
    \midrule
    Cirques ($\sim$4.5~km) & Cirque de Gavarnie & \emph{mainstream} & 0.44 & 24 & 75.6k & 159 & 19 & not on a trail & 17m \\
                           & Cirque d'Estaubé & \emph{known} & 0.67 & 5 & 5.7k & 27 & 5 & not on a trail & 1h08 \\
    \midrule
    Peaks ($\sim$0.4~km)   & Punda Negra / Tallón & \emph{mainstream} & 0.54 & 12 & 15.2k & 33 & 9 & not on a trail & 33m \\
                           & Pico Royo & \emph{partially known} & 0.80 & 0 & -- & 0 & 6 & not on a trail & 37m \\
    \midrule
    Huts ($\sim$2.8~km)    & Refuge des Espuguettes & \emph{mainstream} & 0.55 & 1 & 816 & 8 & 7 & on a national trail + website & 24m \\
                           & Cabane de Labassa & \emph{lesser-known} & 0.90 & 0 & -- & 0 & 1 & on a national trail & 2h29 \\
    \bottomrule
  \end{tabular}
\end{table*}

Despite these first results, our Obscurity Score still has some limitations:

\textbf{Absence as signal}: following the literature, our approach treats low visibility or documentation as evidence that a place is lesser-known \cite{rajaonarivo2023few,pradipta2021journee}, yet in VGI, sparse documentation may instead reflect an inactive mapper community rather than genuine obscurity \cite{haklay2010many}; we currently mitigate this via an initial completeness filter, but selecting areas with an active local mapping community remains essential for the score's trustworthiness.

\textbf{Recognition evolves over time}: tracking edit timestamps lets us observe such shifts, so the score must be treated as a fluid metric rather than a fixed label; publishing a list of lesser-known sites can itself redirect visitor flows and risk new overtourism cases, motivating a future carrying-capacity layer per POI so that promoting hidden gems does not simply displace overcrowding elsewhere. 

\textbf{Obscurity Score versus LLMs}: a further question is how LLMs, which also encode cultural and touristic visibility, compare to our Obscurity Score when prompted for hidden gems; their judgments should be evaluated against local expert assessments alongside our own. Even if LLMs matched or exceeded our approach, our open-data pipeline would remain free and reproducible, motivating hybrid designs that combine both.

\section{Conclusion and Future Work}
\label{sec:conclusion}
In this study, we introduced a generic pipeline that leverages open data to uncover lesser-known mountain POIs through a dedicated scoring system. We demonstrated its potential through preliminary results from an initial case study in the Pyrenees, comparing POIs within the same category to illustrate how the score operates in practice. The expert-in-the-loop calibration itself remains future work, as the collection of validation data is currently ongoing.

Several directions follow: repeating the calibration protocol
across territories and POI types to test weight stability; extending
the same-category comparisons into a full recommendation system
that considers trail distance, elevation gain, trail difficulty, site
quality, and similarity to propose alternatives; and integrating
a carrying-capacity layer so that promoting lesser-known sites
remains conditioned on their ability to host additional visitors.
Ultimately, this work lays the foundation for identifying lesser-known sites, providing a key step toward redistributing tourist
flows and enriching the promotion of local territorial heritage.

\begin{acks}
This research was developed under a joint supervision between the Université de Pau et des Pays
de l’Adour and the Transilvania University of Brașov. This project has received funding from the
European Union’s Horizon Europe research and innovation programme under the Marie
Skłodowska-Curie grant agreement No 101126643. Funded by the European Union. Views and
opinions expressed are, however, those of the author(s) only and do not necessarily reflect those of
the European Union or the Research Executive Agency. Neither the European Union nor the
granting authority can be held responsible for them.
\end{acks}

\textbf{GenAI Usage Disclosure}
In accordance with ACM policy, we disclose limited use of generative AI tools during manuscript preparation. Specifically, Claude.ai was employed exclusively for language refinement, including grammar corrections, sentence restructuring, and stylistic polishing. All technical content, experiments, analysis, and conclusions were developed solely by the authors; the use of generative AI did not contribute to the creation of any novel ideas, data analysis, or experimental design.

\bibliographystyle{ACM-Reference-Format}
\bibliography{main}

\end{document}